\documentclass[%
 reprint,
 amsmath,amssymb,
 aps,
 prl,
]{revtex4-1}

\usepackage{graphicx}
\usepackage{dcolumn}
\usepackage{bm}

\begin{document}

\preprint{APS/123-QED}

\title{Primordial Sex Facilitates the Emergence of Evolution}

\author{Sam Sinai$^{1,2,*}$, Jason Olejarz$^{1,*}$, Iulia A. Neagu$^{1,3}$, and Martin A. Nowak$^{1,2,4,\dagger}$}
\affiliation{$^1$Program for Evolutionary Dynamics, Harvard University, Cambridge, Massachusetts 02138, USA\\
             $^2$Department of Organismic and Evolutionary Biology, Harvard University, Cambridge, Massachusetts 02138, USA\\
             $^3$Department of Physics, Harvard University, Cambridge, Massachusetts 02138, USA\\
             $^4$Department of Mathematics, Harvard University, Cambridge, Massachusetts 02138, USA}

\thanks{S. S. and J. O. contributed equally to this work.}
\email{$^\dagger$ martin\_nowak@harvard.edu}





\date{\today}
\begin{abstract}

Compartments are ubiquitous throughout biology, yet their importance stretches back to the origin of cells. In the context of origin of life, we assume that a protocell, a compartment enclosing functional components, requires $N$ components to be evolvable. We calculate the timescale in which a minimal evolvable protocell is produced. We show that when protocells fuse and share information, the time to produce an evolvable protocell scales algebraically in $N$, in contrast to an exponential scaling in the absence of fusion. We discuss the implications of this result for origins of life, as well as other biological processes. 

\end{abstract}

\pacs{Valid PACS appear here}
\maketitle

A defining characteristic of living organisms is their ability to replicate and evolve ~\cite{NowakBook}.  A major objective of research on the origin of life is therefore to find plausible chemical systems that are capable of self-replication.  The ``RNA world hypothesis'' is a leading framework encompassing theories about the role of RNA in the origin of life. It postulates that RNA or a similar bio-polymer, being both an information-carrying molecule, as well as an enzymatic one, must have played a central role in initiating self-replication~\cite{Crick,Orgel,Woese}. But formidable difficulties remain for developing this narrative into a complete and rigorous theory of the origin of life. Both theoretical and experimental investigations show that well-mixed populations of RNA or similar bio-polymers often suffer from calamitous pitfalls, including the error catastrophe for replicases ~\cite{Eigen71} and parasitism for cooperative enzymes ~\cite{bansho2016host,Nowak2014,Szathmary2006,Hogeweg}. Moreover, the complexity of long RNA sequences that could serve as efficient catalysts creates a challenge for explaining their spontaneous prebiotic synthesis ~\cite{Bernhardt_2012}. Indeed, despite decades of efforts in prebiotic chemistry (and some exciting progress, e.g. ~\cite{Attwater2013,Joyce2009}), building efficient, stable, and prebiotically plausible replicases (sometimes called the holy grail of the RNA world) has remained a challenge \cite{Pross_2013,RNAworldreview}.  

In modern cells, lipid membranes compartmentalize information-carrying and enzymatic molecules akin to those sought after by RNA world researchers. Hence, at some point in the development of life, either before, during, or after the emergence of self-replicating genetic elements, such compartmentalization must have occurred. There is evidence in support of the prebiotic availability of lipid membranes. It has been shown that amphiphilic molecules, like simple fatty acids that are building blocks for the lipid membrane, can be produced in a prebiotically plausible manner \cite{McCollom}.  Alternatively, lipids could have been imported to earth by chondrite meteorites \cite{Yuen84,Yuen79,Deamer}.  Hence, such molecules were likely abundantly present on the prebiotic Earth \cite{Lancet2001, Lane2012, Luisi, Paleos, deamer1986}.  These molecules are able to spontaneously assemble into lipid vesicles in aqueous conditions \cite{yamam,deamer1986}, forming compartments, which in this context are known as \emph{protocells}.

Protocells alleviate some of the pitfalls that an impede the transition from prelife to life. The contents of protocells are held near each other and share the same fate. This results in increased interactions within the protocell and decreased interactions with the outside environment. It also means that the protocell can house a segmented genome, i.e. the information within the protocell need not be stored in one contiguous polymer. It can also dampen the effects of side reactions for any auto-catalytic cycles that may be required to start and maintain a metabolism \cite{SzathmaryChapter}. Protocells can also divide into new protocells that inherit parts of their contents \cite{Markvoort2007,Fanelli_2008}. These properties of protocells enable them to help in selection for cooperative polymers, in particular replicases \cite{bansho2016host,Nowak2014,Nowak2013,Szostak2009,Szostak2012,Szostak2013,Hanczyc}. In addition to enclosing information and dividing, protocells are able to merge, thereby sharing their contents \cite{KrapivskyRednerBook,Bernstein,SzathmarySex,SzathmaryChapter}. In biology, sharing information content between two individuals is considered a defining property of sex. 

The implications of this information-sharing ability among protocells, which is a form of ``primordial sex", have not received much attention. For reasons outlined in the rest of this study, we suggest that the ability for these compartments to merge categorically changes the time required to produce an evolvable protocell. Hence, we propose that early presence of membranes, possibly even before the advent of replication, could have vastly improved the chances of producing complicated cells by luck. In such cases it would not be unreasonable to assume that the starting set of molecules from which an evolvable cell emerges could be large. Almost no origin of life models operate on this assumption, because they consider it a probabilistic miracle. 

To test this hypothesis, we investigate a simple first-passage process \cite{RednerBook,Chou_2014}.  We assume that in order to be evolvable, a protocell needs to contain a certain number, $N$, of component types (i.e., distinct molecules of various complexity) \cite{Szathmary2006, Szathmary1987,Lancet2000, Ganti75, Kauffman86,Vaidya}. In early life, these could be molecules as simple as ions, activated monomers, molecules that stabilize the membrane, or more complicated polymers, like oligo-peptides, and even elementary ribozymes and simple unlinked genes \cite{Szostak2009, Szostak2012, Joyce2009, Szostak2008,Fishkis,Chen2012,SzathmaryChapter,Black_2013,bar1997dynamics,vasas2012evolution}.  More precisely, the target set should result in an auto-catalytic network that results in a evolvable cell with non-negligible probability. Such a scheme has been proposed since Oparin, and has been defended more recently \cite{vasas2012evolution}. We term the smallest set of necessary and sufficient components from which an evolvable protocell can be made a \emph{minimal evolvable protocell}.

\begin{figure}
\centering
\includegraphics*[width=0.45\textwidth]{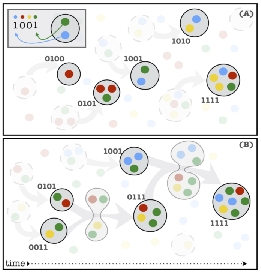}
\caption{Merging occurs between randomly assembled protocells.  (A)  Each color (and a ``1'' bit at each corresponding position on a protocell's representative binary string) indicates presence of one of the four components needed for the protocell to be evolvable (here, $N=4$).  Randomly assembled lipid membranes form around the components.  (B)  Whenever two protocells merge, they share their contents.  Sharing of contents is computed as a bitwise OR operation between each of the two parent strings of length $N$.}
\label{fig:cells}
\end{figure}

We can accordingly represent the functional (or genetic) content of each protocell as a binary string of length $N$. For simplicity, we ignore the redundancy (or dose) of each component in the protocell, and are only concerned with each component's presence. If a protocell contains a particular component $i$, then the string will have a value of 1 at the $i^\mathrm{th}$ position and 0 otherwise. Whenever a protocell randomly assembles, we assume that it contains each of the $N$ component types independently (components do not compete for positions) with probability $p$. I.e. protocell assembly uniformly samples each type (with sufficient abundance) from the environment with probability $p$. Whenever two protocells merge, the value of the resulting string at every position $i$ is simply determined by a bitwise OR operation on the $i^\mathrm{th}$ bits of the two parent protocells (i.e. if either of the original cells contain a component, the resulting cell will also contain it). This is shown schematically in Figure \ref{fig:cells}.

The dynamical process is as follows.  On the first step, the \emph{accumulator}---the object of our attention---consists of a randomly assembled protocell.  If less than $N$ components are enclosed, then one of two things can happen:  With probability $\delta$, the accumulator loses its contents, and on the second step, the accumulator consists of a new randomly assembled protocell, with the accumulation process starting over.  The accumulator can lose its contents if, for example, its membrane's integrity is lost, it is infected by a parasite, or it divides, and the parameter $\delta$ accounts for all such possibilities.  Or with probability $1-\delta$, on the second step, the accumulator merges with a randomly assembled protocell from the environment, possibly gaining additional components.  In this case, if the accumulator still has less than $N$ components after merging, then one of two things can happen:  With probability $\delta$, the accumulator loses its contents, and on the third step, the accumulator consists of a new randomly assembled protocell, with the accumulation process starting over.  Or with probability $1-\delta$, on the third step, the accumulator merges with another randomly assembled protocell from the environment, possibly gaining additional components.  This process continues until the accumulator gains all $N$ components necessary for evolvability.  The total number of steps (or time units), $Z$, needed to gain all $N$ components is equal to the total number of random assembly and merging events in the accumulation process.

The time, $Z$, needed to form a minimal evolvable protocell is thus a random variable that depends on the particular accumulator being tracked.  If we track many such accumulators, then what is the mean first-passage time, $E[Z]$, for an accumulator to achieve all $N$ components necessary for evolvability?

Begin by considering the simple case $\delta=1$ (no merging occurs).  If the accumulator consists of a randomly assembled protocell that has all $N$ components, then the minimal evolvable protocell has been achieved.  But if there are less than $N$ components, then the accumulator is reset without merging.  Thus, the expected number of such random assembly events required to accumulate all $N$ components necessary for evolvability, $\mathrm{E}_{\delta=1}[Z]$, grows exponentially with $N$, i.e.,
\begin{equation*}
\mathrm{E}_{\delta=1}[Z] = \left(\frac{1}{p}\right)^N
\end{equation*}
For large values of $N$, the spontaneous generation of a minimal evolvable protocell would be a probabilistic miracle.  We now focus our attention on understanding how $E[Z]$ grows with $N$ when $0 < \delta <1$. 

In what follows, it is convenient to use the parameter $q \equiv 1-p$.  Denote by $S(q,\delta,N)$ the probability that, starting from a randomly assembled protocell, the accumulator achieves all $N$ components before being reset.  We determine $S(q,\delta,N)$ as follows. First, assume that there is no death of the accumulator. Then $1-q^z$ is the probability that, after $z$ steps, the accumulator has achieved a component.  Therefore, $1-(1-q^z)^N$ is the probability that the accumulator has not achieved all $N$ components after $z$ steps.  It follows that $(1-q^z)^N-(1-q^{z-1})^N$ is the probability that the accumulator achieves all $N$ components in exactly $z$ steps.  Then, considering death of the accumulator, since the probability that the accumulator survives for $z$ steps without being reset is simply $(1-\delta)^{z-1}$ , we have

\begin{equation*}
S(q,\delta,N)=\sum_{z=1}^\infty(1-\delta)^{z-1}\left[\left(1-q^z\right)^N-\left(1-q^{z-1}\right)^N\right]
\end{equation*}
This can be simplified as
\begin{equation}
S(q,\delta,N)=\frac{\delta}{1-\delta}\sum_{z=1}^\infty(1-\delta)^z\left(1-q^z\right)^N
\label{eqn:S}
\end{equation}
Denote by $T(z;q,\delta,N)$ the probability mass function for the number of steps, $z$, needed for the accumulator to gain all $N$ components (i.e., reach its target) when starting from a randomly assembled protocell, given that all $N$ components are accumulated before being reset.  We have
\begin{equation}
T(z;q,\delta,N) = \frac{(1-\delta)^{z-1}\left[\left(1-q^z\right)^N-\left(1-q^{z-1}\right)^N\right]}{S(q,\delta,N)}
\label{eqn:T}
\end{equation}
Denote by $R(z;q,\delta,N)$ the probability mass function for the number of steps, $z$, taken before the accumulator is reset when starting from a randomly assembled protocell, given that the accumulator is reset before gaining all $N$ components.  We have
\begin{equation}
R(z;q,\delta,N) = \frac{\delta(1-\delta)^{z-1}\left[1-\left(1-q^z\right)^N\right]}{1-S(q,\delta,N)}
\label{eqn:R}
\end{equation}
In what follows, we omit explicitly writing the functional dependencies on $q$, $\delta$, and $N$ for notational convenience.

For all $0 < \delta < 1$, the mean first-passage time, $E[Z]$, needed to form a minimal evolvable protocell is calculated directly from
\begin{equation}
\mathrm{E}[Z] = \frac{\sum_{z=1}^\infty z \left[ ST(z)+(1-S)R(z) \right]}{S}
\label{eqn:EZ1}
\end{equation}
Substituting Eqs. \eqref{eqn:S}, \eqref{eqn:T}, and \eqref{eqn:R} into Eq. \eqref{eqn:EZ1} and simplifying, we obtain
\begin{equation}
\mathrm{E}[Z] = \frac{1}{S\delta}-\frac{1-\delta}{\delta}
\label{eqn:EZ_all_N}
\end{equation}

To extract the large-$N$ behavior of $\mathrm{E}[Z]$ from Eq. \eqref{eqn:EZ_all_N}, we simplify the summation in Eq. \eqref{eqn:S} for large $N$ using the following procedure.  For a smooth function $f(x)$, we use the notation $f^{(i)}(x)=d^if(x)/dx^i$.  We can express an integration of $f^{(i)}(x)$ with respect to $x$ from $0$ to $\infty$ as
\begin{equation*}
\int_0^\infty dx \; f^{(i)}(x) = \frac{1}{N} \sum_{z=0}^\infty \int_0^1 dy \; f^{(i)}\left(\frac{z+y}{N}\right)
\end{equation*}
Next, we write a Taylor expansion of $f^{(i)}((z+y)/N)$ in powers of $y/N$ and perform the integration over $y$.  We have
\begin{equation}
\int_0^\infty dx \, f^{(i)}(x) = \sum_{m=0}^\infty \frac{1}{(m+1)!N^m} \left[ \frac{1}{N} \sum_{z=0}^\infty f^{(i+m)}\left(\frac{z}{N}\right) \right]
\label{eqn:identity}
\end{equation}

Substituting Eq. \eqref{eqn:identity} into Eq. \eqref{eqn:S} to express the summation as an integration, substituting the integral form of the Beta function, $B(x,y)=\int_0^1 \mathop{dt} t^{x-1} (1-t)^{y-1}$, and using $\sim$ to denote asymptotic equivalence as $N\rightarrow\infty$, we obtain
\begin{equation}
S \sim \frac{-\delta}{(1-\delta) \log (q)} B \left(\frac{\log(1-\delta)}{\log (q)},N+1\right)
\label{eqn:S_asymptotic}
\end{equation}
Substituting Eq. \eqref{eqn:S_asymptotic} into Eq. \eqref{eqn:EZ_all_N}, expressing the Beta function using Gamma components, $B(x,y)=\Gamma(x)\Gamma(y)/\Gamma(x+y)$, using Stirling's formula for the Gamma function, $\Gamma(x) \sim x^xe^{-x}\sqrt{2\pi/x}$, and simplifying for large $N$, we find that $\mathrm{E}[Z]$ grows asymptotically as
\begin{equation}
\begin{aligned}
\mathrm{E}[Z] \sim \alpha N^k,
\label{eqn:EZ}
\end{aligned}
\end{equation}
where
\begin{equation*}
\alpha = \frac{- (1-\delta)\log (1-p)}{\delta^2\Gamma(k)}
\end{equation*}
and
\begin{equation*}
k = \frac{\log(1-\delta)}{\log(1-p)}.
\end{equation*}

\begin{figure}
\centering
\includegraphics*[width=0.45\textwidth]{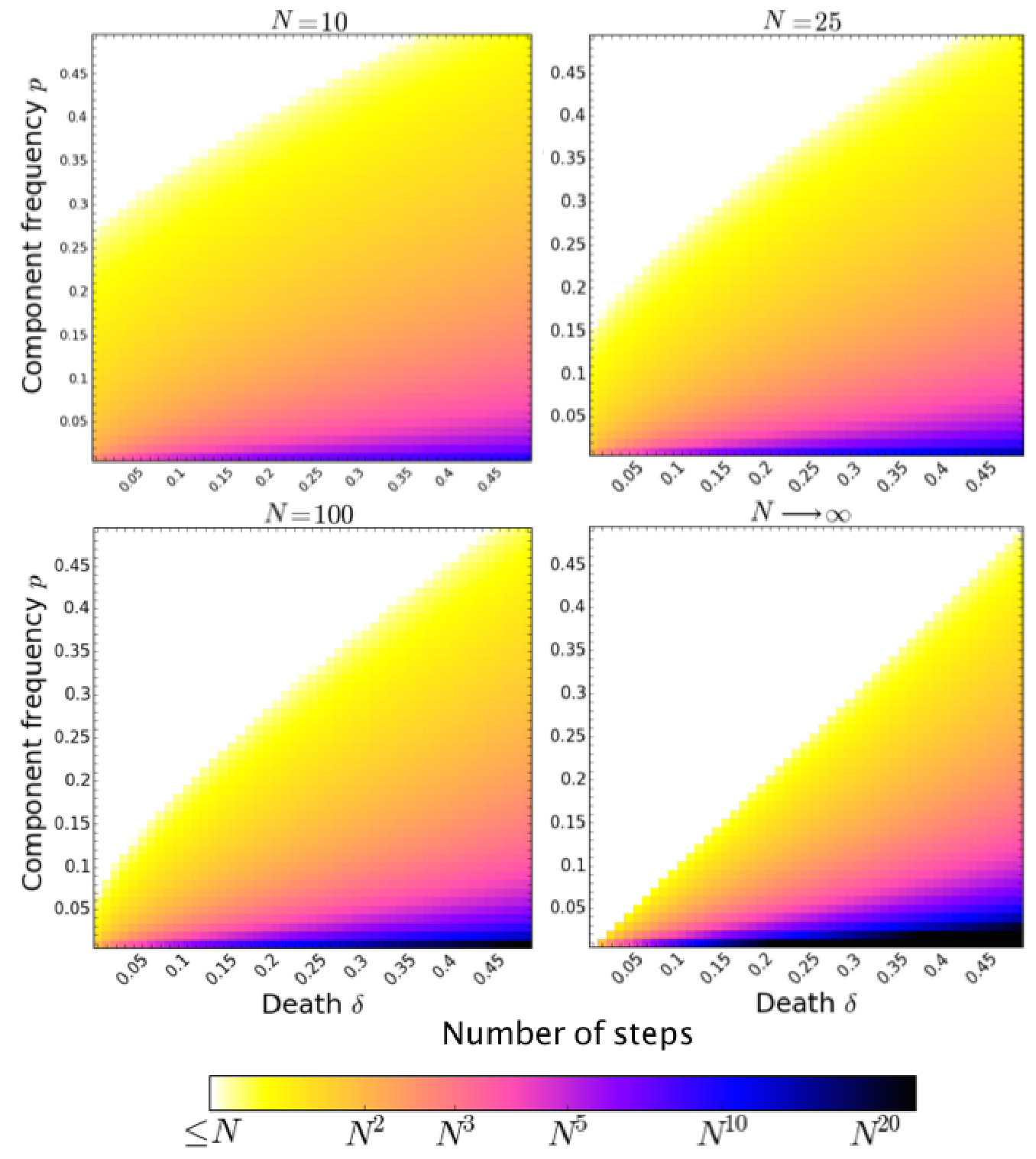}
\caption{Minimal evolvable protocells are achieved in polynomial time for the vast majority of parameter space.  For $N=10$, $N=25$, and $N=100$, we perform Monte Carlo simulations of the accumulation of components, and we plot $N^{\log_N(E[Z])}$ as a function of $p$ and $\delta$.  For $N\rightarrow\infty$, we plot $N^k$ as a function of $p$ and $\delta$.}
\label{fig:N}
\end{figure}

The time complexity of concurrence of components for the problem of abiogenesis is thus fundamentally altered:  For any slight amount of merging, i.e., for any value $0 <\delta<1$, $E[Z]$ grows algebraically with $N$.  Intriguingly, for many values of $p$ and $\delta$, $E[Z]$ grows only as a small power of $N$, and for many other values of $p$ and $\delta$, $E[Z]$ grows only sublinearly with $N$ (Figure \ref{fig:N}).

\begin{figure}
\centering
\includegraphics*[width=0.45\textwidth]{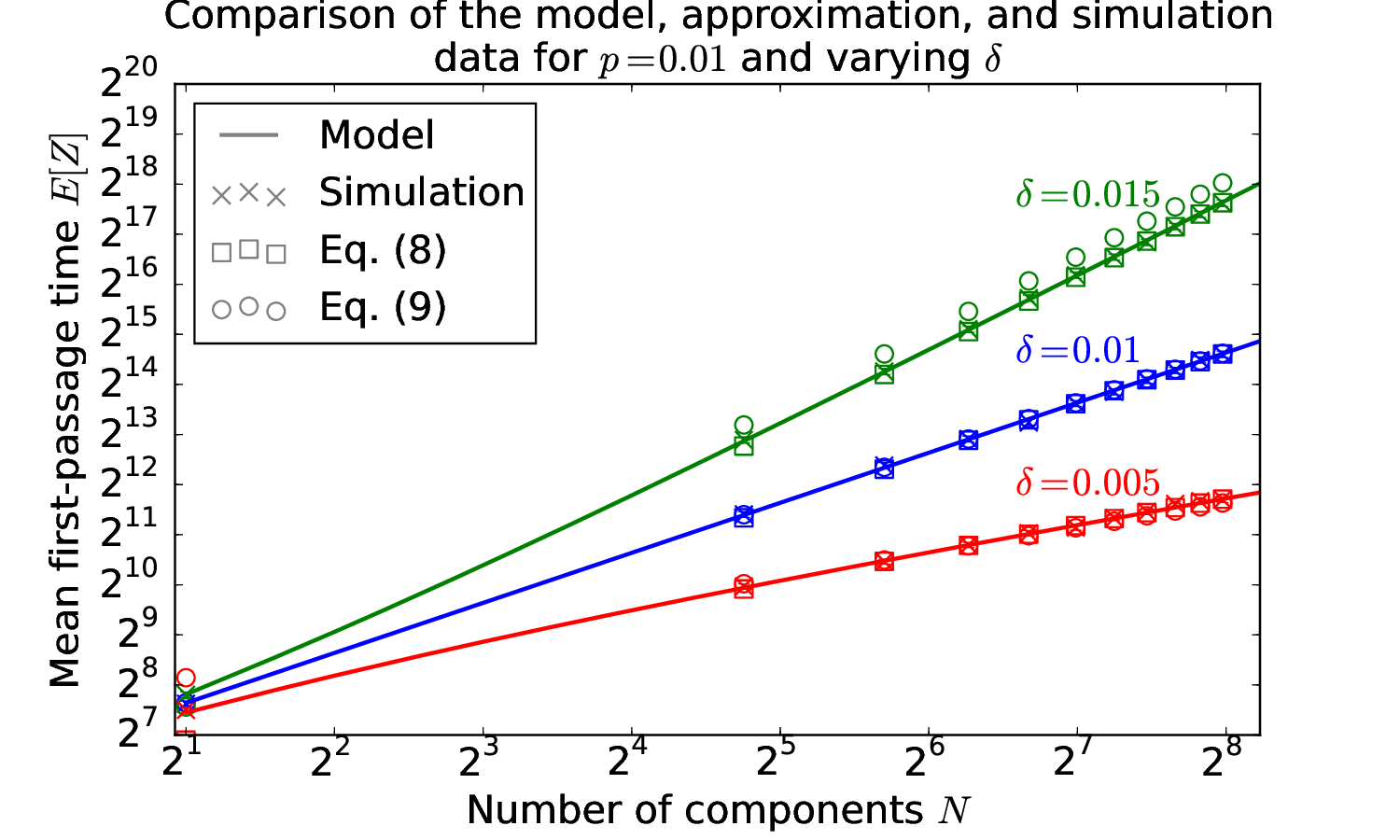}
\caption{For $p=0.01$ and $\delta=0.005$, $\delta=0.01$, and $\delta=0.015$, we plot the exact form of $E[Z]$ for any value of $N$ given by Eq. \eqref{eqn:EZ_all_N} (solid lines), $E[Z]$ measured using a Monte Carlo simulation of the accumulation of components ($\times$), the asymptotically exact form of $E[Z]$ given by Eq. \eqref{eqn:EZ} ($\square$), and the approximation for $E[Z]$ given by Eq. \eqref{eqn:EZ_approx} ($\circ$).}
\label{fig:curves}
\end{figure}

For the particular case in which $\delta \ll 1$, $p \ll 1$, and $\delta$ is not too large relative to $p$, Eq. \eqref{eqn:EZ} admits a simple approximation:
\begin{equation}
\mathrm{E}[Z] \approx \frac{1}{\delta} N^{\delta/p}
\label{eqn:EZ_approx}
\end{equation}
The exact form of $E[Z]$ for all values of $N$ given by Eq. \eqref{eqn:EZ_all_N}, $E[Z]$ measured using a Monte Carlo simulation of the accumulation of components, the exact asymptotic result for $E[Z]$ given by Eq. \eqref{eqn:EZ}, and the approximation for $E[Z]$ given by Eq. \eqref{eqn:EZ_approx} are plotted in Figure \ref{fig:curves} for several values of $p$, $\delta$, and $N$.

For the particular case $\delta=0$, the accumulator begins as a randomly assembled protocell, and it is never reset.  For this case, the mean first-passage time, $\mathrm{E}_{\delta=0}[Z]$, grows logarithmically with $N$, i.e. \footnote{A reader with interest in algorithms may recognize this result as the expected number of levels in a skip list with N elements, where 1-p is the probability of each element's appearance in the next level of the skip list \cite{skiplist,skiplist2}.},
\begin{equation*}
\mathrm{E}_{\delta=0}[Z] = \sum_{i=1}^N {N \choose i}\frac{(-1)^{i+1}}{1-(1-p)^i} \sim \frac{\log(N)}{-\log(1-p)}
\end{equation*}

Also of interest for the biologically realistic case $0 < \delta < 1$ is the probability mass function, $P(Z=z)$, for the number of steps needed to achieve a minimal evolvable protocell.  $P(Z=z)$ is given by
\begin{equation}
P(Z=z) = S \sum_{i=1}^z (1-S)^{i-1} \sum_{\sum_{j=1}^i z_j = z} T(z_1) \prod_{j \neq 1} R(z_j)
\label{eqn:variance_explicit}
\end{equation}

If $N$ is small, then there is typically a small number of resets before the accumulator gains all components, which corresponds to each $z_j$ being comparable in magnitude to $z$ in the summations in Eq. \eqref{eqn:variance_explicit}. But if $N$ is large, then there is typically a large number of resets before the accumulator gains all components, which corresponds to having $z_j \ll z$ for all $j$ in the summations in Eq. \eqref{eqn:variance_explicit}. In this case, the total number of steps, $Z$, is the sum of many independent and identically distributed random variables. 

To provide a sense of how well of an estimator $E[Z]$ is for the variable $Z$ we look at its concentration $\tilde{Z}=Z/E[Z]$. Denote $\mu$ as the average number of steps before an accumulator resets given that the accumulator resets before gaining all N components.  Denote $\sigma^2$ as the variance in the number of steps before an accumulator resets given that the accumulator resets before gaining all N components. We have $\mu = \sum_{z=1}^\infty zR(z) \sim 1/\delta$ and $\sigma^2 = \left(\sum_{z=1}^\infty z^2R(z)\right) - \mu^2 \sim (1-\delta)/\delta^2$.  Since both $\mu$ and $\sigma^2$ are finite, the central limit theorem enables a simplification of Eq. \eqref{eqn:variance_explicit} for large values of $N$: we obtain the probability density function for $\tilde{Z}$:
\begin{equation}
P(\tilde{Z}=\tilde{z}) \sim \mathrm{E}[Z] \sum_{i=1}^\infty \frac{S(1-S)^i}{\sqrt{2\pi i \sigma^2}}\exp \left( \frac{-(\mathrm{E}[Z]\tilde{z}-i\mu)^2}{2i\sigma^2} \right) \\
\label{eqn:P_ztilde}
\end{equation}
The moments of $\tilde{Z}$ are directly computed from Equation ~\eqref{eqn:P_ztilde}:
\begin{equation*}
\mathrm{E}[\tilde{Z}^m] \sim \int_0^\infty d\tilde{z} \; \tilde{z}^m \; P(\tilde{Z}=\tilde{z}) \sim m!
\end{equation*}
We immediately see that $Z$ is exponentially distributed about $E[Z]$:
\begin{equation*}
P(Z=z) \sim \frac{1}{\alpha N^k} \exp \left( -\frac{z}{\alpha N^k} \right)
\end{equation*}
For large $N$, the natural production of minimal evolvable protocells via random assembly and repeated fusion is therefore simply a Poisson process.

It is noteworthy that $E[Z]$, Eq. \eqref{eqn:EZ_all_N}, provides an upper bound on the time to construct a minimal evolvable protocell for many natural variations of this process. There are many ways in which the first-passage time can be shortened. For instance, the expected time to reach the target set of N components is reduced if cells divide (and retain some components) instead of losing all components through death. Redundancy in components, where a protocell might have one or more backup copies of each component, can have a similar effect. Moreover, our simple model is specified by only three parameters.  Our model is therefore robust for exploring the time complexity of myriad of compartmentalization scenarios by simply tuning the values of $p$, $\delta$, and $N$.

Doing so will help in understanding several biological questions, and relates our work to other studies that are interested in the timescale of evolutionary events. For instance, Wilf and Ewens \cite{wilf2010there} arrive at exactly the same formula for $\delta=0$ when looking for the time it takes for evolution (on a smooth landscape with a single peak, hence $\delta=0$) to reach a target set of genes. This analysis is also favorable to viewing sex as the default biological state. Computational analyses of sex suggest that it makes evolutionary search over a landscape more efficient\cite{Krish, SexLivnat,livnat2016sex}. Our analysis adds that this advantage could be present, and aid, in starting cellular replication itself. While biologists have considered the possibility of early sex before \cite{Bernstein,Haldane}, it was soon observed that parasitism could be a serious problem \cite{SzathmaryChapter}. However, our exact asymptotic analysis instead suggests that sex is a good strategy, even in the presence of parasites.

Oceanic currents in early earth could have brought together primitive protocells with disparate components, which subsequently merged and eventually spawned an evolvable protocell.  In this scenario, protocell formation, convection, and merging act as a necessary bridge between physically and chemically heterogeneous prebiotic environments for biological construction.  Indeed, there is exciting, ongoing experimental work on creating ``self-sustaining'' protocells, which can divide and subsequently restore their viable composition via fusion for a few generations \cite{kurihara2015recursive}.  

Our mathematical model is similarly well suited for investigating the biological activation of modern viruses.  In particular, our model captures a process known as \emph{multiplicity reactivation}. In this process multiple non-functional, mutant viruses of the same strain combine, thereby ``covering'' each other's loss-of-function mutations and producing a functioning virus. Our analysis readily provides the number of such viral particles required (in expectation) that would re-activate a virus. In a similar scenario, in multi-compartment viruses, multiple distinct components need to co-infect the same host in order to produce a new virion.  In many plant viruses, such as the genus \emph{Tymovirus}, the infection occurs when two or more functionally distinct virions infect the same host \cite{EGsmith,PlantVir}.  The occurrence of this type of combinatorial reproduction in many RNA viruses, which are thought to be ancient, is consistent with the thesis that primordial sex played an integral role in early life \cite{Nasir,Koonin}.

Research into minimal synthetic cells has shown that cells with few hundred genes are able to self-sustain in complex media \cite{Mushegian_1996,Gil_2004,Hutchison_2016}. This suggests that even for low values of $p$, in this case the probability of required genes vs random protein-coding genes, novel self-sustaining cells (and possibly viruses) could be produced, either in lab or in early life, by a feasible number of fusions. The feasibility of finding novel viable combinations through a merging process in lab should also be of help in order to understand the density of viable solutions within the fitness landscape \cite{Krish}.
 
We may never know with certainty what path has resulted in the emergence of life on earth. There are likely many possible paths to evolvability, none of which have been fully delineated to this date. So far, virtually all models of protocells assumed a small initial set size, precisely because co-occurrence of many components together is unlikely. We show that even if the number of required components $N$ is large, there are tenable paths to construct such an assembly. The merging mechanism is not as critical if $N$ is small, but in the presence of merging compartments we are no longer restricted to this scenario. Here, we have devised and analyzed a model that captures a general set of possibilities for an evolvable protocell to emerge. It is noteworthy that our model remains agnostic about whether template-directed replication or metabolism emerged first and it can apply in both scenarios as well as different levels of complexity (from chemicals to enzymes and genes) in the underlying components. 

To the best of our knowledge, our study is the first to provide a rigorous and quantitative blueprint for comparing the plausibility of a subset of paths to life: those that involve compartmentalization. 
 
\begin{acknowledgments}
The authors thank Krishnendu Chatterjee for comments about the manuscript. We also thank Leslie Valiant and Scott Linderman for helpful comments in the initial phases of this project. We thank Robert Israel for pointing us to related literature. We thank Michael Nicholson and Nicolas Fraiman for helpful discussions. We thank Artem Kaznatcheev for a great discussion of our preprint on his blog. This work was supported by the John Templeton Foundation and in part by a grant from B. Wu and Eric Larson.
\end{acknowledgments}


\bibliography{Sinai_Olejarz_Neagu_Nowak_v2}

\begin{thebibliography}{64}%
\makeatletter
\providecommand \@ifxundefined [1]{%
 \@ifx{#1\undefined}
}%
\providecommand \@ifnum [1]{%
 \ifnum #1\expandafter \@firstoftwo
 \else \expandafter \@secondoftwo
 \fi
}%
\providecommand \@ifx [1]{%
 \ifx #1\expandafter \@firstoftwo
 \else \expandafter \@secondoftwo
 \fi
}%
\providecommand \natexlab [1]{#1}%
\providecommand \enquote  [1]{``#1''}%
\providecommand \bibnamefont  [1]{#1}%
\providecommand \bibfnamefont [1]{#1}%
\providecommand \citenamefont [1]{#1}%
\providecommand \href@noop [0]{\@secondoftwo}%
\providecommand \href [0]{\begingroup \@sanitize@url \@href}%
\providecommand \@href[1]{\@@startlink{#1}\@@href}%
\providecommand \@@href[1]{\endgroup#1\@@endlink}%
\providecommand \@sanitize@url [0]{\catcode `\\12\catcode `\$12\catcode
  `\&12\catcode `\#12\catcode `\^12\catcode `\_12\catcode `\%12\relax}%
\providecommand \@@startlink[1]{}%
\providecommand \@@endlink[0]{}%
\providecommand \url  [0]{\begingroup\@sanitize@url \@url }%
\providecommand \@url [1]{\endgroup\@href {#1}{\urlprefix }}%
\providecommand \urlprefix  [0]{URL }%
\providecommand \Eprint [0]{\href }%
\providecommand \doibase [0]{http://dx.doi.org/}%
\providecommand \selectlanguage [0]{\@gobble}%
\providecommand \bibinfo  [0]{\@secondoftwo}%
\providecommand \bibfield  [0]{\@secondoftwo}%
\providecommand \translation [1]{[#1]}%
\providecommand \BibitemOpen [0]{}%
\providecommand \bibitemStop [0]{}%
\providecommand \bibitemNoStop [0]{.\EOS\space}%
\providecommand \EOS [0]{\spacefactor3000\relax}%
\providecommand \BibitemShut  [1]{\csname bibitem#1\endcsname}%
\let\auto@bib@innerbib\@empty
\bibitem [{\citenamefont {Nowak}(2006)}]{NowakBook}%
  \BibitemOpen
  \bibfield  {author} {\bibinfo {author} {\bibfnamefont {M.~A.}\ \bibnamefont
  {Nowak}},\ }\href@noop {} {\emph {\bibinfo {title} {Evolutionary dynamics}}}\
  (\bibinfo  {publisher} {Harvard University Press},\ \bibinfo {year}
  {2006})\BibitemShut {NoStop}%
\bibitem [{\citenamefont {Crick}\ \emph {et~al.}(1968)\citenamefont {Crick}
  \emph {et~al.}}]{Crick}%
  \BibitemOpen
  \bibfield  {author} {\bibinfo {author} {\bibfnamefont {F.~H.}\ \bibnamefont
  {Crick}} \emph {et~al.},\ }\href@noop {} {\bibfield  {journal} {\bibinfo
  {journal} {Journal of molecular biology}\ }\textbf {\bibinfo {volume} {38}},\
  \bibinfo {pages} {367} (\bibinfo {year} {1968})}\BibitemShut {NoStop}%
\bibitem [{\citenamefont {Orgel}(1968)}]{Orgel}%
  \BibitemOpen
  \bibfield  {author} {\bibinfo {author} {\bibfnamefont {L.~E.}\ \bibnamefont
  {Orgel}},\ }\href@noop {} {\bibfield  {journal} {\bibinfo  {journal} {Journal
  of molecular biology}\ }\textbf {\bibinfo {volume} {38}},\ \bibinfo {pages}
  {381} (\bibinfo {year} {1968})}\BibitemShut {NoStop}%
\bibitem [{\citenamefont {Woese}(1967)}]{Woese}%
  \BibitemOpen
  \bibfield  {author} {\bibinfo {author} {\bibfnamefont {C.}~\bibnamefont
  {Woese}},\ }\href@noop {} {\emph {\bibinfo {title} {The genetic code}}}\
  (\bibinfo  {publisher} {Harper and Row, New York},\ \bibinfo {year}
  {1967})\BibitemShut {NoStop}%
\bibitem [{\citenamefont {Eigen}(1971)}]{Eigen71}%
  \BibitemOpen
  \bibfield  {author} {\bibinfo {author} {\bibfnamefont {M.}~\bibnamefont
  {Eigen}},\ }\href@noop {} {\bibfield  {journal} {\bibinfo  {journal}
  {Naturwissenschaften}\ }\textbf {\bibinfo {volume} {58}},\ \bibinfo {pages}
  {465} (\bibinfo {year} {1971})}\BibitemShut {NoStop}%
\bibitem [{\citenamefont {Bansho}\ \emph {et~al.}(2016)\citenamefont {Bansho},
  \citenamefont {Furubayashi}, \citenamefont {Ichihashi},\ and\ \citenamefont
  {Yomo}}]{bansho2016host}%
  \BibitemOpen
  \bibfield  {author} {\bibinfo {author} {\bibfnamefont {Y.}~\bibnamefont
  {Bansho}}, \bibinfo {author} {\bibfnamefont {T.}~\bibnamefont {Furubayashi}},
  \bibinfo {author} {\bibfnamefont {N.}~\bibnamefont {Ichihashi}}, \ and\
  \bibinfo {author} {\bibfnamefont {T.}~\bibnamefont {Yomo}},\ }\href@noop {}
  {\bibfield  {journal} {\bibinfo  {journal} {Proceedings of the National
  Academy of Sciences}\ ,\ \bibinfo {pages} {201524404}} (\bibinfo {year}
  {2016})}\BibitemShut {NoStop}%
\bibitem [{\citenamefont {Markvoort}\ \emph {et~al.}(2014)\citenamefont
  {Markvoort}, \citenamefont {Sinai},\ and\ \citenamefont {Nowak}}]{Nowak2014}%
  \BibitemOpen
  \bibfield  {author} {\bibinfo {author} {\bibfnamefont {A.~J.}\ \bibnamefont
  {Markvoort}}, \bibinfo {author} {\bibfnamefont {S.}~\bibnamefont {Sinai}}, \
  and\ \bibinfo {author} {\bibfnamefont {M.~A.}\ \bibnamefont {Nowak}},\
  }\href@noop {} {\bibfield  {journal} {\bibinfo  {journal} {Journal of
  theoretical biology}\ }\textbf {\bibinfo {volume} {357}},\ \bibinfo {pages}
  {123} (\bibinfo {year} {2014})}\BibitemShut {NoStop}%
\bibitem [{\citenamefont {Fontanari}\ \emph {et~al.}(2006)\citenamefont
  {Fontanari}, \citenamefont {Santos},\ and\ \citenamefont
  {Szathm{\'a}ry}}]{Szathmary2006}%
  \BibitemOpen
  \bibfield  {author} {\bibinfo {author} {\bibfnamefont {J.~F.}\ \bibnamefont
  {Fontanari}}, \bibinfo {author} {\bibfnamefont {M.}~\bibnamefont {Santos}}, \
  and\ \bibinfo {author} {\bibfnamefont {E.}~\bibnamefont {Szathm{\'a}ry}},\
  }\href@noop {} {\bibfield  {journal} {\bibinfo  {journal} {Journal of
  theoretical biology}\ }\textbf {\bibinfo {volume} {239}},\ \bibinfo {pages}
  {247} (\bibinfo {year} {2006})}\BibitemShut {NoStop}%
\bibitem [{\citenamefont {Hogeweg}\ and\ \citenamefont
  {Takeuchi}(2003)}]{Hogeweg}%
  \BibitemOpen
  \bibfield  {author} {\bibinfo {author} {\bibfnamefont {P.}~\bibnamefont
  {Hogeweg}}\ and\ \bibinfo {author} {\bibfnamefont {N.}~\bibnamefont
  {Takeuchi}},\ }\href@noop {} {\bibfield  {journal} {\bibinfo  {journal}
  {Origins of Life and Evolution of the Biosphere}\ }\textbf {\bibinfo {volume}
  {33}},\ \bibinfo {pages} {375} (\bibinfo {year} {2003})}\BibitemShut
  {NoStop}%
\bibitem [{\citenamefont {Bernhardt}(2012)}]{Bernhardt_2012}%
  \BibitemOpen
  \bibfield  {author} {\bibinfo {author} {\bibfnamefont {H.~S.}\ \bibnamefont
  {Bernhardt}},\ }\href {\doibase 10.1186/1745-6150-7-23} {\bibfield  {journal}
  {\bibinfo  {journal} {Biology Direct}\ }\textbf {\bibinfo {volume} {7}}
  (\bibinfo {year} {2012}),\ 10.1186/1745-6150-7-23}\BibitemShut {NoStop}%
\bibitem [{\citenamefont {Attwater}\ \emph {et~al.}(2013)\citenamefont
  {Attwater}, \citenamefont {Wochner},\ and\ \citenamefont
  {Holliger}}]{Attwater2013}%
  \BibitemOpen
  \bibfield  {author} {\bibinfo {author} {\bibfnamefont {J.}~\bibnamefont
  {Attwater}}, \bibinfo {author} {\bibfnamefont {A.}~\bibnamefont {Wochner}}, \
  and\ \bibinfo {author} {\bibfnamefont {P.}~\bibnamefont {Holliger}},\
  }\href@noop {} {\bibfield  {journal} {\bibinfo  {journal} {Nature chemistry}\
  }\textbf {\bibinfo {volume} {5}},\ \bibinfo {pages} {1011} (\bibinfo {year}
  {2013})}\BibitemShut {NoStop}%
\bibitem [{\citenamefont {Lincoln}\ and\ \citenamefont
  {Joyce}(2009)}]{Joyce2009}%
  \BibitemOpen
  \bibfield  {author} {\bibinfo {author} {\bibfnamefont {T.~A.}\ \bibnamefont
  {Lincoln}}\ and\ \bibinfo {author} {\bibfnamefont {G.~F.}\ \bibnamefont
  {Joyce}},\ }\href@noop {} {\bibfield  {journal} {\bibinfo  {journal}
  {Science}\ }\textbf {\bibinfo {volume} {323}},\ \bibinfo {pages} {1229}
  (\bibinfo {year} {2009})}\BibitemShut {NoStop}%
\bibitem [{\citenamefont {Pross}\ and\ \citenamefont
  {Pascal}(2013)}]{Pross_2013}%
  \BibitemOpen
  \bibfield  {author} {\bibinfo {author} {\bibfnamefont {A.}~\bibnamefont
  {Pross}}\ and\ \bibinfo {author} {\bibfnamefont {R.}~\bibnamefont {Pascal}},\
  }\href {\doibase 10.1098/rsob.120190} {\bibfield  {journal} {\bibinfo
  {journal} {Open Biology}\ }\textbf {\bibinfo {volume} {3}} (\bibinfo {year}
  {2013}),\ 10.1098/rsob.120190}\BibitemShut {NoStop}%
\bibitem [{\citenamefont {Higgs}\ and\ \citenamefont
  {Lehman}(2014)}]{RNAworldreview}%
  \BibitemOpen
  \bibfield  {author} {\bibinfo {author} {\bibfnamefont {P.~G.}\ \bibnamefont
  {Higgs}}\ and\ \bibinfo {author} {\bibfnamefont {N.}~\bibnamefont {Lehman}},\
  }\href@noop {} {\bibfield  {journal} {\bibinfo  {journal} {Nature Reviews
  Genetics}\ } (\bibinfo {year} {2014})}\BibitemShut {NoStop}%
\bibitem [{\citenamefont {McCollom}\ \emph {et~al.}(1999)\citenamefont
  {McCollom}, \citenamefont {Ritter},\ and\ \citenamefont
  {Simoneit}}]{McCollom}%
  \BibitemOpen
  \bibfield  {author} {\bibinfo {author} {\bibfnamefont {T.~M.}\ \bibnamefont
  {McCollom}}, \bibinfo {author} {\bibfnamefont {G.}~\bibnamefont {Ritter}}, \
  and\ \bibinfo {author} {\bibfnamefont {B.~R.}\ \bibnamefont {Simoneit}},\
  }\href@noop {} {\bibfield  {journal} {\bibinfo  {journal} {Origins of Life
  and Evolution of the Biosphere}\ }\textbf {\bibinfo {volume} {29}},\ \bibinfo
  {pages} {153} (\bibinfo {year} {1999})}\BibitemShut {NoStop}%
\bibitem [{\citenamefont {Yuen}\ \emph {et~al.}(1984)\citenamefont {Yuen},
  \citenamefont {Blair}, \citenamefont {Des~Marais},\ and\ \citenamefont
  {Chang}}]{Yuen84}%
  \BibitemOpen
  \bibfield  {author} {\bibinfo {author} {\bibfnamefont {G.}~\bibnamefont
  {Yuen}}, \bibinfo {author} {\bibfnamefont {N.}~\bibnamefont {Blair}},
  \bibinfo {author} {\bibfnamefont {D.~J.}\ \bibnamefont {Des~Marais}}, \ and\
  \bibinfo {author} {\bibfnamefont {S.}~\bibnamefont {Chang}},\ }\href@noop {}
  {\bibfield  {journal} {\bibinfo  {journal} {Nature}\ } (\bibinfo {year}
  {1984})}\BibitemShut {NoStop}%
\bibitem [{\citenamefont {Lawless}\ and\ \citenamefont {Yuen}(1979)}]{Yuen79}%
  \BibitemOpen
  \bibfield  {author} {\bibinfo {author} {\bibfnamefont {J.~G.}\ \bibnamefont
  {Lawless}}\ and\ \bibinfo {author} {\bibfnamefont {G.~U.}\ \bibnamefont
  {Yuen}},\ }\href@noop {} {\bibfield  {journal} {\bibinfo  {journal} {Nature}\
  } (\bibinfo {year} {1979})}\BibitemShut {NoStop}%
\bibitem [{\citenamefont {Deamer}(1985)}]{Deamer}%
  \BibitemOpen
  \bibfield  {author} {\bibinfo {author} {\bibfnamefont {D.~W.}\ \bibnamefont
  {Deamer}},\ }\href@noop {} {\bibfield  {journal} {\bibinfo  {journal}
  {Nature}\ } (\bibinfo {year} {1985})}\BibitemShut {NoStop}%
\bibitem [{\citenamefont {Segr{\'e}}\ \emph {et~al.}(2001)\citenamefont
  {Segr{\'e}}, \citenamefont {Ben-Eli}, \citenamefont {Deamer},\ and\
  \citenamefont {Lancet}}]{Lancet2001}%
  \BibitemOpen
  \bibfield  {author} {\bibinfo {author} {\bibfnamefont {D.}~\bibnamefont
  {Segr{\'e}}}, \bibinfo {author} {\bibfnamefont {D.}~\bibnamefont {Ben-Eli}},
  \bibinfo {author} {\bibfnamefont {D.~W.}\ \bibnamefont {Deamer}}, \ and\
  \bibinfo {author} {\bibfnamefont {D.}~\bibnamefont {Lancet}},\ }\href@noop {}
  {\bibfield  {journal} {\bibinfo  {journal} {Origins of Life and Evolution of
  the Biosphere}\ }\textbf {\bibinfo {volume} {31}},\ \bibinfo {pages} {119}
  (\bibinfo {year} {2001})}\BibitemShut {NoStop}%
\bibitem [{\citenamefont {Lane}\ and\ \citenamefont {Martin}(2012)}]{Lane2012}%
  \BibitemOpen
  \bibfield  {author} {\bibinfo {author} {\bibfnamefont {N.}~\bibnamefont
  {Lane}}\ and\ \bibinfo {author} {\bibfnamefont {W.~F.}\ \bibnamefont
  {Martin}},\ }\href@noop {} {\bibfield  {journal} {\bibinfo  {journal} {Cell}\
  }\textbf {\bibinfo {volume} {151}},\ \bibinfo {pages} {1406} (\bibinfo {year}
  {2012})}\BibitemShut {NoStop}%
\bibitem [{\citenamefont {Luisi}\ \emph {et~al.}(1999)\citenamefont {Luisi},
  \citenamefont {Walde},\ and\ \citenamefont {Oberholzer}}]{Luisi}%
  \BibitemOpen
  \bibfield  {author} {\bibinfo {author} {\bibfnamefont {P.~L.}\ \bibnamefont
  {Luisi}}, \bibinfo {author} {\bibfnamefont {P.}~\bibnamefont {Walde}}, \ and\
  \bibinfo {author} {\bibfnamefont {T.}~\bibnamefont {Oberholzer}},\
  }\href@noop {} {\bibfield  {journal} {\bibinfo  {journal} {Current opinion in
  colloid \& interface science}\ }\textbf {\bibinfo {volume} {4}},\ \bibinfo
  {pages} {33} (\bibinfo {year} {1999})}\BibitemShut {NoStop}%
\bibitem [{\citenamefont {Paleos}(2015)}]{Paleos}%
  \BibitemOpen
  \bibfield  {author} {\bibinfo {author} {\bibfnamefont {C.}~\bibnamefont
  {Paleos}},\ }\href@noop {} {\bibfield  {journal} {\bibinfo  {journal} {Trends
  in Biochemical Sciences}\ }\textbf {\bibinfo {volume} {40}},\ \bibinfo
  {pages} {487} (\bibinfo {year} {2015})}\BibitemShut {NoStop}%
\bibitem [{\citenamefont {Deamer}(1986)}]{deamer1986}%
  \BibitemOpen
  \bibfield  {author} {\bibinfo {author} {\bibfnamefont {D.~W.}\ \bibnamefont
  {Deamer}},\ }\href@noop {} {\bibfield  {journal} {\bibinfo  {journal}
  {Origins of Life and Evolution of the Biosphere}\ }\textbf {\bibinfo {volume}
  {17}},\ \bibinfo {pages} {3} (\bibinfo {year} {1986})}\BibitemShut {NoStop}%
\bibitem [{\citenamefont {Yamamoto}\ \emph {et~al.}(2002)\citenamefont
  {Yamamoto}, \citenamefont {Maruyama},\ and\ \citenamefont {Hyodo}}]{yamam}%
  \BibitemOpen
  \bibfield  {author} {\bibinfo {author} {\bibfnamefont {S.}~\bibnamefont
  {Yamamoto}}, \bibinfo {author} {\bibfnamefont {Y.}~\bibnamefont {Maruyama}},
  \ and\ \bibinfo {author} {\bibfnamefont {S.-a.}\ \bibnamefont {Hyodo}},\
  }\href@noop {} {\bibfield  {journal} {\bibinfo  {journal} {The Journal of
  Chemical Physics}\ }\textbf {\bibinfo {volume} {116}},\ \bibinfo {pages}
  {5842} (\bibinfo {year} {2002})}\BibitemShut {NoStop}%
\bibitem [{\citenamefont {Szathm{\'a}ry}\ \emph {et~al.}(2005)\citenamefont
  {Szathm{\'a}ry}, \citenamefont {Santos},\ and\ \citenamefont
  {Fernando}}]{SzathmaryChapter}%
  \BibitemOpen
  \bibfield  {author} {\bibinfo {author} {\bibfnamefont {E.}~\bibnamefont
  {Szathm{\'a}ry}}, \bibinfo {author} {\bibfnamefont {M.}~\bibnamefont
  {Santos}}, \ and\ \bibinfo {author} {\bibfnamefont {C.}~\bibnamefont
  {Fernando}},\ }in\ \href@noop {} {\emph {\bibinfo {booktitle} {Prebiotic
  Chemistry}}}\ (\bibinfo  {publisher} {Springer},\ \bibinfo {year} {2005})\
  pp.\ \bibinfo {pages} {167--211}\BibitemShut {NoStop}%
\bibitem [{\citenamefont {Markvoort}\ \emph {et~al.}(2007)\citenamefont
  {Markvoort}, \citenamefont {Smeijers}, \citenamefont {Pieterse},
  \citenamefont {van Santen},\ and\ \citenamefont {Hilbers}}]{Markvoort2007}%
  \BibitemOpen
  \bibfield  {author} {\bibinfo {author} {\bibfnamefont {A.}~\bibnamefont
  {Markvoort}}, \bibinfo {author} {\bibfnamefont {A.}~\bibnamefont {Smeijers}},
  \bibinfo {author} {\bibfnamefont {K.}~\bibnamefont {Pieterse}}, \bibinfo
  {author} {\bibfnamefont {R.}~\bibnamefont {van Santen}}, \ and\ \bibinfo
  {author} {\bibfnamefont {P.}~\bibnamefont {Hilbers}},\ }\href@noop {}
  {\bibfield  {journal} {\bibinfo  {journal} {The Journal of Physical Chemistry
  B}\ }\textbf {\bibinfo {volume} {111}},\ \bibinfo {pages} {5719} (\bibinfo
  {year} {2007})}\BibitemShut {NoStop}%
\bibitem [{\citenamefont {Fanelli}\ and\ \citenamefont
  {McKane}(2008)}]{Fanelli_2008}%
  \BibitemOpen
  \bibfield  {author} {\bibinfo {author} {\bibfnamefont {D.}~\bibnamefont
  {Fanelli}}\ and\ \bibinfo {author} {\bibfnamefont {A.~J.}\ \bibnamefont
  {McKane}},\ }\href {\doibase 10.1103/PhysRevE.78.051406} {\bibfield
  {journal} {\bibinfo  {journal} {Physical Review E}\ }\textbf {\bibinfo
  {volume} {78}} (\bibinfo {year} {2008}),\
  10.1103/PhysRevE.78.051406}\BibitemShut {NoStop}%
\bibitem [{\citenamefont {Bianconi}\ \emph {et~al.}(2013)\citenamefont
  {Bianconi}, \citenamefont {Zhao}, \citenamefont {Chen},\ and\ \citenamefont
  {Nowak}}]{Nowak2013}%
  \BibitemOpen
  \bibfield  {author} {\bibinfo {author} {\bibfnamefont {G.}~\bibnamefont
  {Bianconi}}, \bibinfo {author} {\bibfnamefont {K.}~\bibnamefont {Zhao}},
  \bibinfo {author} {\bibfnamefont {I.~A.}\ \bibnamefont {Chen}}, \ and\
  \bibinfo {author} {\bibfnamefont {M.~A.}\ \bibnamefont {Nowak}},\ }\href@noop
  {} {\bibfield  {journal} {\bibinfo  {journal} {PLoS Comput Biol}\ }\textbf
  {\bibinfo {volume} {9}},\ \bibinfo {pages} {e1003051} (\bibinfo {year}
  {2013})}\BibitemShut {NoStop}%
\bibitem [{\citenamefont {Zhu}\ and\ \citenamefont
  {Szostak}(2009)}]{Szostak2009}%
  \BibitemOpen
  \bibfield  {author} {\bibinfo {author} {\bibfnamefont {T.~F.}\ \bibnamefont
  {Zhu}}\ and\ \bibinfo {author} {\bibfnamefont {J.~W.}\ \bibnamefont
  {Szostak}},\ }\href@noop {} {\bibfield  {journal} {\bibinfo  {journal}
  {Journal of the American Chemical Society}\ }\textbf {\bibinfo {volume}
  {131}},\ \bibinfo {pages} {5705} (\bibinfo {year} {2009})}\BibitemShut
  {NoStop}%
\bibitem [{\citenamefont {Budin}\ \emph {et~al.}(2012)\citenamefont {Budin},
  \citenamefont {Debnath},\ and\ \citenamefont {Szostak}}]{Szostak2012}%
  \BibitemOpen
  \bibfield  {author} {\bibinfo {author} {\bibfnamefont {I.}~\bibnamefont
  {Budin}}, \bibinfo {author} {\bibfnamefont {A.}~\bibnamefont {Debnath}}, \
  and\ \bibinfo {author} {\bibfnamefont {J.~W.}\ \bibnamefont {Szostak}},\
  }\href@noop {} {\bibfield  {journal} {\bibinfo  {journal} {Journal of the
  American Chemical Society}\ }\textbf {\bibinfo {volume} {134}},\ \bibinfo
  {pages} {20812} (\bibinfo {year} {2012})}\BibitemShut {NoStop}%
\bibitem [{\citenamefont {Adamala}\ and\ \citenamefont
  {Szostak}(2013)}]{Szostak2013}%
  \BibitemOpen
  \bibfield  {author} {\bibinfo {author} {\bibfnamefont {K.}~\bibnamefont
  {Adamala}}\ and\ \bibinfo {author} {\bibfnamefont {J.~W.}\ \bibnamefont
  {Szostak}},\ }\href@noop {} {\bibfield  {journal} {\bibinfo  {journal}
  {Science}\ }\textbf {\bibinfo {volume} {342}},\ \bibinfo {pages} {1098}
  (\bibinfo {year} {2013})}\BibitemShut {NoStop}%
\bibitem [{\citenamefont {Hanczyc}\ \emph {et~al.}(2003)\citenamefont
  {Hanczyc}, \citenamefont {Fujikawa},\ and\ \citenamefont
  {Szostak}}]{Hanczyc}%
  \BibitemOpen
  \bibfield  {author} {\bibinfo {author} {\bibfnamefont {M.~M.}\ \bibnamefont
  {Hanczyc}}, \bibinfo {author} {\bibfnamefont {S.~M.}\ \bibnamefont
  {Fujikawa}}, \ and\ \bibinfo {author} {\bibfnamefont {J.~W.}\ \bibnamefont
  {Szostak}},\ }\href@noop {} {\bibfield  {journal} {\bibinfo  {journal}
  {Science}\ }\textbf {\bibinfo {volume} {302}},\ \bibinfo {pages} {618}
  (\bibinfo {year} {2003})}\BibitemShut {NoStop}%
\bibitem [{\citenamefont {Krapivsky}\ \emph {et~al.}(2010)\citenamefont
  {Krapivsky}, \citenamefont {Redner},\ and\ \citenamefont
  {Ben-Naim}}]{KrapivskyRednerBook}%
  \BibitemOpen
  \bibfield  {author} {\bibinfo {author} {\bibfnamefont {P.~L.}\ \bibnamefont
  {Krapivsky}}, \bibinfo {author} {\bibfnamefont {S.}~\bibnamefont {Redner}}, \
  and\ \bibinfo {author} {\bibfnamefont {E.}~\bibnamefont {Ben-Naim}},\
  }\href@noop {} {\emph {\bibinfo {title} {A Kinetic View of Statistical
  Physics}}}\ (\bibinfo  {publisher} {Cambridge University Press},\ \bibinfo
  {year} {2010})\BibitemShut {NoStop}%
\bibitem [{\citenamefont {Bernstein}\ \emph {et~al.}(1984)\citenamefont
  {Bernstein}, \citenamefont {Byerly}, \citenamefont {Hopf},\ and\
  \citenamefont {Michod}}]{Bernstein}%
  \BibitemOpen
  \bibfield  {author} {\bibinfo {author} {\bibfnamefont {H.}~\bibnamefont
  {Bernstein}}, \bibinfo {author} {\bibfnamefont {H.~C.}\ \bibnamefont
  {Byerly}}, \bibinfo {author} {\bibfnamefont {F.~A.}\ \bibnamefont {Hopf}}, \
  and\ \bibinfo {author} {\bibfnamefont {R.~E.}\ \bibnamefont {Michod}},\
  }\href@noop {} {\bibfield  {journal} {\bibinfo  {journal} {Journal of
  Theoretical Biology}\ }\textbf {\bibinfo {volume} {110}},\ \bibinfo {pages}
  {323} (\bibinfo {year} {1984})}\BibitemShut {NoStop}%
\bibitem [{\citenamefont {Santos}\ \emph {et~al.}(2003)\citenamefont {Santos},
  \citenamefont {Zintzaras},\ and\ \citenamefont
  {Szathm{\'a}ry}}]{SzathmarySex}%
  \BibitemOpen
  \bibfield  {author} {\bibinfo {author} {\bibfnamefont {M.}~\bibnamefont
  {Santos}}, \bibinfo {author} {\bibfnamefont {E.}~\bibnamefont {Zintzaras}}, \
  and\ \bibinfo {author} {\bibfnamefont {E.}~\bibnamefont {Szathm{\'a}ry}},\
  }\href@noop {} {\bibfield  {journal} {\bibinfo  {journal} {Origins of Life
  and Evolution of the Biosphere}\ }\textbf {\bibinfo {volume} {33}},\ \bibinfo
  {pages} {405} (\bibinfo {year} {2003})}\BibitemShut {NoStop}%
\bibitem [{\citenamefont {Redner}(2001)}]{RednerBook}%
  \BibitemOpen
  \bibfield  {author} {\bibinfo {author} {\bibfnamefont {S.}~\bibnamefont
  {Redner}},\ }\href@noop {} {\emph {\bibinfo {title} {A Guide to First-Passage
  Processes}}}\ (\bibinfo  {publisher} {Cambridge University Press},\ \bibinfo
  {year} {2001})\BibitemShut {NoStop}%
\bibitem [{\citenamefont {Chou}\ and\ \citenamefont
  {D'Orsogna}(2014)}]{Chou_2014}%
  \BibitemOpen
  \bibfield  {author} {\bibinfo {author} {\bibfnamefont {T.}~\bibnamefont
  {Chou}}\ and\ \bibinfo {author} {\bibfnamefont {M.~R.}\ \bibnamefont
  {D'Orsogna}},\ }\enquote {\bibinfo {title} {First passage problems in
  biology},}\ \ (\bibinfo  {publisher} {World Scientific},\ \bibinfo {year}
  {2014})\ Chap.~\bibinfo {chapter} {13}, pp.\ \bibinfo {pages}
  {306--345}\BibitemShut {NoStop}%
\bibitem [{\citenamefont {Szathm{\'a}ry}\ and\ \citenamefont
  {Demeter}(1987)}]{Szathmary1987}%
  \BibitemOpen
  \bibfield  {author} {\bibinfo {author} {\bibfnamefont {E.}~\bibnamefont
  {Szathm{\'a}ry}}\ and\ \bibinfo {author} {\bibfnamefont {L.}~\bibnamefont
  {Demeter}},\ }\href@noop {} {\bibfield  {journal} {\bibinfo  {journal}
  {Journal of theoretical biology}\ }\textbf {\bibinfo {volume} {128}},\
  \bibinfo {pages} {463} (\bibinfo {year} {1987})}\BibitemShut {NoStop}%
\bibitem [{\citenamefont {Segr{\'e}}\ \emph {et~al.}(2000)\citenamefont
  {Segr{\'e}}, \citenamefont {Ben-Eli},\ and\ \citenamefont
  {Lancet}}]{Lancet2000}%
  \BibitemOpen
  \bibfield  {author} {\bibinfo {author} {\bibfnamefont {D.}~\bibnamefont
  {Segr{\'e}}}, \bibinfo {author} {\bibfnamefont {D.}~\bibnamefont {Ben-Eli}},
  \ and\ \bibinfo {author} {\bibfnamefont {D.}~\bibnamefont {Lancet}},\
  }\href@noop {} {\bibfield  {journal} {\bibinfo  {journal} {Proceedings of the
  National Academy of Sciences}\ }\textbf {\bibinfo {volume} {97}},\ \bibinfo
  {pages} {4112} (\bibinfo {year} {2000})}\BibitemShut {NoStop}%
\bibitem [{\citenamefont {G{\'a}nti}(1975)}]{Ganti75}%
  \BibitemOpen
  \bibfield  {author} {\bibinfo {author} {\bibfnamefont {T.}~\bibnamefont
  {G{\'a}nti}},\ }\href@noop {} {\bibfield  {journal} {\bibinfo  {journal}
  {BioSystems}\ }\textbf {\bibinfo {volume} {7}},\ \bibinfo {pages} {15}
  (\bibinfo {year} {1975})}\BibitemShut {NoStop}%
\bibitem [{\citenamefont {Kauffman}(1986)}]{Kauffman86}%
  \BibitemOpen
  \bibfield  {author} {\bibinfo {author} {\bibfnamefont {S.~A.}\ \bibnamefont
  {Kauffman}},\ }\href@noop {} {\bibfield  {journal} {\bibinfo  {journal}
  {Journal of theoretical biology}\ }\textbf {\bibinfo {volume} {119}},\
  \bibinfo {pages} {1} (\bibinfo {year} {1986})}\BibitemShut {NoStop}%
\bibitem [{\citenamefont {Vaidya}\ \emph {et~al.}(2012)\citenamefont {Vaidya},
  \citenamefont {Manapat}, \citenamefont {Chen}, \citenamefont {Xulvi-Brunet},
  \citenamefont {Hayden},\ and\ \citenamefont {Lehman}}]{Vaidya}%
  \BibitemOpen
  \bibfield  {author} {\bibinfo {author} {\bibfnamefont {N.}~\bibnamefont
  {Vaidya}}, \bibinfo {author} {\bibfnamefont {M.~L.}\ \bibnamefont {Manapat}},
  \bibinfo {author} {\bibfnamefont {I.~A.}\ \bibnamefont {Chen}}, \bibinfo
  {author} {\bibfnamefont {R.}~\bibnamefont {Xulvi-Brunet}}, \bibinfo {author}
  {\bibfnamefont {E.~J.}\ \bibnamefont {Hayden}}, \ and\ \bibinfo {author}
  {\bibfnamefont {N.}~\bibnamefont {Lehman}},\ }\href@noop {} {\bibfield
  {journal} {\bibinfo  {journal} {Nature}\ }\textbf {\bibinfo {volume} {491}},\
  \bibinfo {pages} {72} (\bibinfo {year} {2012})}\BibitemShut {NoStop}%
\bibitem [{\citenamefont {Mansy}\ \emph {et~al.}(2008)\citenamefont {Mansy},
  \citenamefont {Schrum}, \citenamefont {Krishnamurthy}, \citenamefont
  {Tob{\'e}}, \citenamefont {Treco},\ and\ \citenamefont
  {Szostak}}]{Szostak2008}%
  \BibitemOpen
  \bibfield  {author} {\bibinfo {author} {\bibfnamefont {S.~S.}\ \bibnamefont
  {Mansy}}, \bibinfo {author} {\bibfnamefont {J.~P.}\ \bibnamefont {Schrum}},
  \bibinfo {author} {\bibfnamefont {M.}~\bibnamefont {Krishnamurthy}}, \bibinfo
  {author} {\bibfnamefont {S.}~\bibnamefont {Tob{\'e}}}, \bibinfo {author}
  {\bibfnamefont {D.~A.}\ \bibnamefont {Treco}}, \ and\ \bibinfo {author}
  {\bibfnamefont {J.~W.}\ \bibnamefont {Szostak}},\ }\href@noop {} {\bibfield
  {journal} {\bibinfo  {journal} {Nature}\ }\textbf {\bibinfo {volume} {454}},\
  \bibinfo {pages} {122} (\bibinfo {year} {2008})}\BibitemShut {NoStop}%
\bibitem [{\citenamefont {Fishkis}(2007)}]{Fishkis}%
  \BibitemOpen
  \bibfield  {author} {\bibinfo {author} {\bibfnamefont {M.}~\bibnamefont
  {Fishkis}},\ }\href@noop {} {\bibfield  {journal} {\bibinfo  {journal}
  {Origins of Life and Evolution of Biospheres}\ }\textbf {\bibinfo {volume}
  {37}},\ \bibinfo {pages} {537} (\bibinfo {year} {2007})}\BibitemShut
  {NoStop}%
\bibitem [{\citenamefont {Chen}\ and\ \citenamefont {Nowak}(2012)}]{Chen2012}%
  \BibitemOpen
  \bibfield  {author} {\bibinfo {author} {\bibfnamefont {I.~A.}\ \bibnamefont
  {Chen}}\ and\ \bibinfo {author} {\bibfnamefont {M.~A.}\ \bibnamefont
  {Nowak}},\ }\href@noop {} {\bibfield  {journal} {\bibinfo  {journal}
  {Accounts of chemical research}\ }\textbf {\bibinfo {volume} {45}},\ \bibinfo
  {pages} {2088} (\bibinfo {year} {2012})}\BibitemShut {NoStop}%
\bibitem [{\citenamefont {Black}\ \emph {et~al.}(2013)\citenamefont {Black},
  \citenamefont {Blosser}, \citenamefont {Stottrup}, \citenamefont {Tavakley},
  \citenamefont {Deamer},\ and\ \citenamefont {Keller}}]{Black_2013}%
  \BibitemOpen
  \bibfield  {author} {\bibinfo {author} {\bibfnamefont {R.~A.}\ \bibnamefont
  {Black}}, \bibinfo {author} {\bibfnamefont {M.~C.}\ \bibnamefont {Blosser}},
  \bibinfo {author} {\bibfnamefont {B.~L.}\ \bibnamefont {Stottrup}}, \bibinfo
  {author} {\bibfnamefont {R.}~\bibnamefont {Tavakley}}, \bibinfo {author}
  {\bibfnamefont {D.~W.}\ \bibnamefont {Deamer}}, \ and\ \bibinfo {author}
  {\bibfnamefont {S.~L.}\ \bibnamefont {Keller}},\ }\href {\doibase
  10.1073/pnas.1300963110} {\bibfield  {journal} {\bibinfo  {journal}
  {Proceedings of the National Academy of Sciences of the United States of
  America}\ }\textbf {\bibinfo {volume} {110}},\ \bibinfo {pages} {13272}
  (\bibinfo {year} {2013})}\BibitemShut {NoStop}%
\bibitem [{\citenamefont {Bar-Yam}(1997)}]{bar1997dynamics}%
  \BibitemOpen
  \bibfield  {author} {\bibinfo {author} {\bibfnamefont {Y.}~\bibnamefont
  {Bar-Yam}},\ }\href@noop {} {\emph {\bibinfo {title} {Dynamics of complex
  systems}}},\ Vol.\ \bibinfo {volume} {213}\ (\bibinfo  {publisher}
  {Addison-Wesley Reading, MA},\ \bibinfo {year} {1997})\BibitemShut {NoStop}%
\bibitem [{\citenamefont {Vasas}\ \emph {et~al.}(2012)\citenamefont {Vasas},
  \citenamefont {Fernando}, \citenamefont {Santos}, \citenamefont {Kauffman},\
  and\ \citenamefont {Szathm{\'a}ry}}]{vasas2012evolution}%
  \BibitemOpen
  \bibfield  {author} {\bibinfo {author} {\bibfnamefont {V.}~\bibnamefont
  {Vasas}}, \bibinfo {author} {\bibfnamefont {C.}~\bibnamefont {Fernando}},
  \bibinfo {author} {\bibfnamefont {M.}~\bibnamefont {Santos}}, \bibinfo
  {author} {\bibfnamefont {S.}~\bibnamefont {Kauffman}}, \ and\ \bibinfo
  {author} {\bibfnamefont {E.}~\bibnamefont {Szathm{\'a}ry}},\ }\href@noop {}
  {\bibfield  {journal} {\bibinfo  {journal} {Biology Direct}\ }\textbf
  {\bibinfo {volume} {7}},\ \bibinfo {pages} {1} (\bibinfo {year}
  {2012})}\BibitemShut {NoStop}%
\bibitem [{Note1()}]{Note1}%
  \BibitemOpen
  \bibinfo {note} {A reader with interest in algorithms may recognize this
  result as the expected number of levels in a skip list with N elements, where
  1-p is the probability of each element's appearance in the next level of the
  skip list \cite {skiplist,skiplist2}.}\BibitemShut {Stop}%
\bibitem [{\citenamefont {Wilf}\ and\ \citenamefont
  {Ewens}(2010)}]{wilf2010there}%
  \BibitemOpen
  \bibfield  {author} {\bibinfo {author} {\bibfnamefont {H.~S.}\ \bibnamefont
  {Wilf}}\ and\ \bibinfo {author} {\bibfnamefont {W.~J.}\ \bibnamefont
  {Ewens}},\ }\href@noop {} {\bibfield  {journal} {\bibinfo  {journal}
  {Proceedings of the National Academy of Sciences}\ }\textbf {\bibinfo
  {volume} {107}},\ \bibinfo {pages} {22454} (\bibinfo {year}
  {2010})}\BibitemShut {NoStop}%
\bibitem [{\citenamefont {Chatterjee}\ \emph {et~al.}(2014)\citenamefont
  {Chatterjee}, \citenamefont {Pavlogiannis}, \citenamefont {Adlam},\ and\
  \citenamefont {Nowak}}]{Krish}%
  \BibitemOpen
  \bibfield  {author} {\bibinfo {author} {\bibfnamefont {K.}~\bibnamefont
  {Chatterjee}}, \bibinfo {author} {\bibfnamefont {A.}~\bibnamefont
  {Pavlogiannis}}, \bibinfo {author} {\bibfnamefont {B.}~\bibnamefont {Adlam}},
  \ and\ \bibinfo {author} {\bibfnamefont {M.~A.}\ \bibnamefont {Nowak}},\
  }\href@noop {} {\bibfield  {journal} {\bibinfo  {journal} {PLoS Comput Biol}\
  }\textbf {\bibinfo {volume} {10}},\ \bibinfo {pages} {e1003818} (\bibinfo
  {year} {2014})}\BibitemShut {NoStop}%
\bibitem [{\citenamefont {Livnat}\ \emph {et~al.}(2008)\citenamefont {Livnat},
  \citenamefont {Papadimitriou}, \citenamefont {Dushoff},\ and\ \citenamefont
  {Feldman}}]{SexLivnat}%
  \BibitemOpen
  \bibfield  {author} {\bibinfo {author} {\bibfnamefont {A.}~\bibnamefont
  {Livnat}}, \bibinfo {author} {\bibfnamefont {C.}~\bibnamefont
  {Papadimitriou}}, \bibinfo {author} {\bibfnamefont {J.}~\bibnamefont
  {Dushoff}}, \ and\ \bibinfo {author} {\bibfnamefont {M.~W.}\ \bibnamefont
  {Feldman}},\ }\href@noop {} {\bibfield  {journal} {\bibinfo  {journal}
  {Proceedings of the National Academy of Sciences}\ }\textbf {\bibinfo
  {volume} {105}},\ \bibinfo {pages} {19803} (\bibinfo {year}
  {2008})}\BibitemShut {NoStop}%
\bibitem [{\citenamefont {Livnat}\ and\ \citenamefont
  {Papadimitriou}(2016)}]{livnat2016sex}%
  \BibitemOpen
  \bibfield  {author} {\bibinfo {author} {\bibfnamefont {A.}~\bibnamefont
  {Livnat}}\ and\ \bibinfo {author} {\bibfnamefont {C.}~\bibnamefont
  {Papadimitriou}},\ }\href@noop {} {\bibfield  {journal} {\bibinfo  {journal}
  {Communications of the ACM}\ }\textbf {\bibinfo {volume} {59}},\ \bibinfo
  {pages} {84} (\bibinfo {year} {2016})}\BibitemShut {NoStop}%
\bibitem [{\citenamefont {Haldane}(1929)}]{Haldane}%
  \BibitemOpen
  \bibfield  {author} {\bibinfo {author} {\bibfnamefont {J.~B.~S.}\
  \bibnamefont {Haldane}},\ }\href@noop {} {\bibfield  {journal} {\bibinfo
  {journal} {Rationalist Annual}\ }\textbf {\bibinfo {volume} {148}},\ \bibinfo
  {pages} {3} (\bibinfo {year} {1929})}\BibitemShut {NoStop}%
\bibitem [{\citenamefont {Kurihara}\ \emph {et~al.}(2015)\citenamefont
  {Kurihara}, \citenamefont {Okura}, \citenamefont {Matsuo}, \citenamefont
  {Toyota}, \citenamefont {Suzuki},\ and\ \citenamefont
  {Sugawara}}]{kurihara2015recursive}%
  \BibitemOpen
  \bibfield  {author} {\bibinfo {author} {\bibfnamefont {K.}~\bibnamefont
  {Kurihara}}, \bibinfo {author} {\bibfnamefont {Y.}~\bibnamefont {Okura}},
  \bibinfo {author} {\bibfnamefont {M.}~\bibnamefont {Matsuo}}, \bibinfo
  {author} {\bibfnamefont {T.}~\bibnamefont {Toyota}}, \bibinfo {author}
  {\bibfnamefont {K.}~\bibnamefont {Suzuki}}, \ and\ \bibinfo {author}
  {\bibfnamefont {T.}~\bibnamefont {Sugawara}},\ }\href@noop {} {\bibfield
  {journal} {\bibinfo  {journal} {Nature communications}\ }\textbf {\bibinfo
  {volume} {6}} (\bibinfo {year} {2015})}\BibitemShut {NoStop}%
\bibitem [{\citenamefont {Smith}(1989)}]{EGsmith}%
  \BibitemOpen
  \bibfield  {author} {\bibinfo {author} {\bibfnamefont {J.~M.}\ \bibnamefont
  {Smith}},\ }\href@noop {} {\emph {\bibinfo {title} {Evolutionary
  genetics.}}}\ (\bibinfo  {publisher} {Oxford University Press},\ \bibinfo
  {year} {1989})\BibitemShut {NoStop}%
\bibitem [{\citenamefont {Rao}(2006)}]{PlantVir}%
  \BibitemOpen
  \bibfield  {author} {\bibinfo {author} {\bibfnamefont {A.}~\bibnamefont
  {Rao}},\ }\href@noop {} {\bibfield  {journal} {\bibinfo  {journal} {Annu.
  Rev. Phytopathol.}\ }\textbf {\bibinfo {volume} {44}},\ \bibinfo {pages} {61}
  (\bibinfo {year} {2006})}\BibitemShut {NoStop}%
\bibitem [{\citenamefont {Nasir}\ and\ \citenamefont
  {Caetano-Anoll{\'e}s}(2015)}]{Nasir}%
  \BibitemOpen
  \bibfield  {author} {\bibinfo {author} {\bibfnamefont {A.}~\bibnamefont
  {Nasir}}\ and\ \bibinfo {author} {\bibfnamefont {G.}~\bibnamefont
  {Caetano-Anoll{\'e}s}},\ }\href@noop {} {\bibfield  {journal} {\bibinfo
  {journal} {Science Advances}\ }\textbf {\bibinfo {volume} {1}},\ \bibinfo
  {pages} {e1500527} (\bibinfo {year} {2015})}\BibitemShut {NoStop}%
\bibitem [{\citenamefont {Koonin}\ \emph {et~al.}(2006)\citenamefont {Koonin},
  \citenamefont {Senkevich},\ and\ \citenamefont {Dolja}}]{Koonin}%
  \BibitemOpen
  \bibfield  {author} {\bibinfo {author} {\bibfnamefont {E.~V.}\ \bibnamefont
  {Koonin}}, \bibinfo {author} {\bibfnamefont {T.~G.}\ \bibnamefont
  {Senkevich}}, \ and\ \bibinfo {author} {\bibfnamefont {V.~V.}\ \bibnamefont
  {Dolja}},\ }\href@noop {} {\bibfield  {journal} {\bibinfo  {journal} {Biol
  Direct}\ }\textbf {\bibinfo {volume} {1}},\ \bibinfo {pages} {29} (\bibinfo
  {year} {2006})}\BibitemShut {NoStop}%
\bibitem [{\citenamefont {Mushegian}\ and\ \citenamefont
  {Koonin}(1996)}]{Mushegian_1996}%
  \BibitemOpen
  \bibfield  {author} {\bibinfo {author} {\bibfnamefont {A.~R.}\ \bibnamefont
  {Mushegian}}\ and\ \bibinfo {author} {\bibfnamefont {E.~V.}\ \bibnamefont
  {Koonin}},\ }\href {\doibase 10.1073/pnas.93.19.10268} {\bibfield  {journal}
  {\bibinfo  {journal} {Proceedings of the National Academy of Sciences of the
  United States of America}\ }\textbf {\bibinfo {volume} {93}},\ \bibinfo
  {pages} {10268} (\bibinfo {year} {1996})}\BibitemShut {NoStop}%
\bibitem [{\citenamefont {Gil}\ \emph {et~al.}(2004)\citenamefont {Gil},
  \citenamefont {Silva}, \citenamefont {Pereto},\ and\ \citenamefont
  {Moya}}]{Gil_2004}%
  \BibitemOpen
  \bibfield  {author} {\bibinfo {author} {\bibfnamefont {R.}~\bibnamefont
  {Gil}}, \bibinfo {author} {\bibfnamefont {F.~J.}\ \bibnamefont {Silva}},
  \bibinfo {author} {\bibfnamefont {J.}~\bibnamefont {Pereto}}, \ and\ \bibinfo
  {author} {\bibfnamefont {A.}~\bibnamefont {Moya}},\ }\href {\doibase
  10.1128/MMBR.68.3.518-537.2004} {\bibfield  {journal} {\bibinfo  {journal}
  {Microbiology and Molecular Biology Reviews}\ }\textbf {\bibinfo {volume}
  {68}},\ \bibinfo {pages} {518} (\bibinfo {year} {2004})}\BibitemShut
  {NoStop}%
\bibitem [{\citenamefont {Hutchison~III}\ \emph {et~al.}(2016)\citenamefont
  {Hutchison~III}, \citenamefont {Chuang}, \citenamefont {Noskov},
  \citenamefont {Assad-Garcia}, \citenamefont {Deerinck}, \citenamefont
  {Ellisman}, \citenamefont {Gill}, \citenamefont {Kannan}, \citenamefont
  {Karas}, \citenamefont {Ma}, \citenamefont {Pelletier}, \citenamefont {Qi},
  \citenamefont {Richter}, \citenamefont {Strychalski}, \citenamefont {Sun},
  \citenamefont {Suzuki}, \citenamefont {Tsvetanova}, \citenamefont {Wise},
  \citenamefont {Smith}, \citenamefont {Glass}, \citenamefont {Merryman},
  \citenamefont {Gibson},\ and\ \citenamefont {Venter}}]{Hutchison_2016}%
  \BibitemOpen
  \bibfield  {author} {\bibinfo {author} {\bibfnamefont {C.~A.}\ \bibnamefont
  {Hutchison~III}}, \bibinfo {author} {\bibfnamefont {R.-Y.}\ \bibnamefont
  {Chuang}}, \bibinfo {author} {\bibfnamefont {V.~N.}\ \bibnamefont {Noskov}},
  \bibinfo {author} {\bibfnamefont {N.}~\bibnamefont {Assad-Garcia}}, \bibinfo
  {author} {\bibfnamefont {T.~J.}\ \bibnamefont {Deerinck}}, \bibinfo {author}
  {\bibfnamefont {M.~H.}\ \bibnamefont {Ellisman}}, \bibinfo {author}
  {\bibfnamefont {J.}~\bibnamefont {Gill}}, \bibinfo {author} {\bibfnamefont
  {K.}~\bibnamefont {Kannan}}, \bibinfo {author} {\bibfnamefont {B.~J.}\
  \bibnamefont {Karas}}, \bibinfo {author} {\bibfnamefont {L.}~\bibnamefont
  {Ma}}, \bibinfo {author} {\bibfnamefont {J.~F.}\ \bibnamefont {Pelletier}},
  \bibinfo {author} {\bibfnamefont {Z.-Q.}\ \bibnamefont {Qi}}, \bibinfo
  {author} {\bibfnamefont {A.}~\bibnamefont {Richter}}, \bibinfo {author}
  {\bibfnamefont {E.~A.}\ \bibnamefont {Strychalski}}, \bibinfo {author}
  {\bibfnamefont {L.}~\bibnamefont {Sun}}, \bibinfo {author} {\bibfnamefont
  {Y.}~\bibnamefont {Suzuki}}, \bibinfo {author} {\bibfnamefont
  {B.}~\bibnamefont {Tsvetanova}}, \bibinfo {author} {\bibfnamefont {K.~S.}\
  \bibnamefont {Wise}}, \bibinfo {author} {\bibfnamefont {H.~O.}\ \bibnamefont
  {Smith}}, \bibinfo {author} {\bibfnamefont {J.~I.}\ \bibnamefont {Glass}},
  \bibinfo {author} {\bibfnamefont {C.}~\bibnamefont {Merryman}}, \bibinfo
  {author} {\bibfnamefont {D.~G.}\ \bibnamefont {Gibson}}, \ and\ \bibinfo
  {author} {\bibfnamefont {J.~C.}\ \bibnamefont {Venter}},\ }\href {\doibase
  10.1126/science.aad6253} {\bibfield  {journal} {\bibinfo  {journal}
  {Science}\ }\textbf {\bibinfo {volume} {351}} (\bibinfo {year} {2016}),\
  10.1126/science.aad6253}\BibitemShut {NoStop}%
\bibitem [{\citenamefont {Pugh}(1990)}]{skiplist}%
  \BibitemOpen
  \bibfield  {author} {\bibinfo {author} {\bibfnamefont {W.}~\bibnamefont
  {Pugh}},\ }\href@noop {} {\bibfield  {journal} {\bibinfo  {journal}
  {Communications of the ACM}\ }\textbf {\bibinfo {volume} {33}},\ \bibinfo
  {pages} {668} (\bibinfo {year} {1990})}\BibitemShut {NoStop}%
\bibitem [{\citenamefont {Kirschenhofer}\ \emph {et~al.}(1995)\citenamefont
  {Kirschenhofer}, \citenamefont {Mart{\'\i}nez},\ and\ \citenamefont
  {Prodinger}}]{skiplist2}%
  \BibitemOpen
  \bibfield  {author} {\bibinfo {author} {\bibfnamefont {P.}~\bibnamefont
  {Kirschenhofer}}, \bibinfo {author} {\bibfnamefont {C.}~\bibnamefont
  {Mart{\'\i}nez}}, \ and\ \bibinfo {author} {\bibfnamefont {H.}~\bibnamefont
  {Prodinger}},\ }\href@noop {} {\bibfield  {journal} {\bibinfo  {journal}
  {Theoretical Computer Science}\ }\textbf {\bibinfo {volume} {144}},\ \bibinfo
  {pages} {199} (\bibinfo {year} {1995})}\BibitemShut {NoStop}%
\end{thebibliography}%

\end{document}